\documentclass[prl,twocolumn,amsmath,superscriptaddress,longbibliography]{revtex4-1}
\usepackage{graphicx,subfigure,epstopdf}
\usepackage{bm}% bold maths

\usepackage[colorlinks=true, letterpaper=true, pdfstartview=FitV,
  linkcolor=blue, citecolor=blue, urlcolor=blue]{hyperref}

\begin{document}

\title{Chiral anomaly in Weyl semimetals from spin current injection}

\author{Yang Gao}

\affiliation{Department of Physics, University of Science and Technology of China,
  Hefei, Anhui 230026, China}
\affiliation{ICQD, Hefei National Laboratory for Physical Sciences at Microscale, University of Science and Technology of China, Hefei, Anhui 230026, China}

\date{\today}

\begin{abstract}
Weyl semimetals are well-known for hosting topologically protected linear band crossings, serving as the analog of the relativistic Weyl Fermions in the condensed matter context. Such analogy persists deeply, allowing the existence of the chiral anomaly under parallel electric and magnetic field in Weyl semimetals. Different from such picture, here we show that, a unique mechanism of the chiral anomaly exists in Weyl semimetals by injecting a spin current with parallel spin polarization and flow direction. The existence of such a chiral anomaly is protected by the topological feature that each Weyl cone can also be a source or drain of the spin field in the momentum space. It leads to measurable experimental signals, such as an electric charge current parallel with an applied magnetic field in the absence of the electric field, and a sharp peak at certain resonant frequency in the injection current in achiral Weyl semimetals through the circular photogalvanic effect. Our work shows that the topological implication of Weyl semimetals goes beyond the link with relativistic Weyl Fermions, and offers a promising scenario to examine the interplay between topology and spin. 
\end{abstract}

\maketitle

%Symmetry and conservation laws are bound by Noether's theorem and of fundamental importance in physics. However, for topological particles, a classical symmetry in the Lagrangian can be broken upon qunatization, a phenomenon referred to as the quantum anomaly.

As a defining property of the relativistic Weyl Fermion, the chiral anomaly refers to the unexpected breaking of the chiral symmetry upon quantizing the Lagrangian~\cite{Adler1969,Bell1969}. In condensed matter physics, the Weyl semimetal offers an ideal platform to examine the chiral anomaly~\cite{Nielsen1983,Aji2012, Zyuzin2012,Son2013,Burkov2018}. The isolated Weyl points with linear dispersion appear in pairs in the Brillouin zone, each of which is chiral, acting as a source or drain of the Berry curvature flux. Such chiral fluxes allow the pumping of electrons from one Weyl cone to its conjugate partner upon parallel electric and magnetic field~\cite{Son2013}, causing the chiral anomaly. 
%The resulting chiral polarization is thus the manifestation of the chiral anomaly.
 As a feature that profoundly demonstrates the topology of the Weyl semimetal, the chiral anomaly has been paid intensive attention to in recent years. It not only leads to intriguing observable phenomena, such as the large negative magnetoresistance~\cite{Son2013,Huang2015,Shekhar2015,Jia2016,Zhang2016,Armitage2018} and magneto-optical Kerr effect~\cite{Zhang2017,Parent2020}, but also inspires further exploration of the topology of Weyl semimetals, through, e.g., the mirror anomaly~\cite{Burkov2018b,Nandy2019}, and the chiral anomaly from inhomogeneous strain~\cite{Liu2013,Cortijo2015,Pikulin2016,Cortijo2016}, based on the similarity of the Weyl point with its relativistic counterpart.

In this work, by exploring the coupling between the topology and spin, we demonstrate a distinct type of the chiral anomaly in Weyl semimetals with no relativistic analog. Especially, by activating the spin degree of freedom of the Weyl Fermion through spin-current injection, electrons can be pumped from one Weyl cone to its conjugate partner. Such behavior can be traced back to the topological feature that a Weyl cone is also a source or drain of the spin field, in addition to the Berry curvature. Since the pumping of electron is determined by the flux of spin field under the spin current injection, it is then protected by such topological feature, and generates a chiral polarization when balanced by the inter-Weyl-node scattering.

This exotic type of chiral anomaly reveals the deep interplay between tooplogy and spin. It can be readily probed in experiments. For example, together with the topological chiral magnetic effect, it can generate a current parallel with the external magnetic field. Moreover, based on the requirement of the circular photogalvanic effect, a sharp peak of the injection current can emerge in an achiral Weyl semimetal at an appropriate resonant light frequency, in response to circularly polarized lights.

\begin{figure}[t]
  \centering
  \includegraphics[width=\columnwidth]{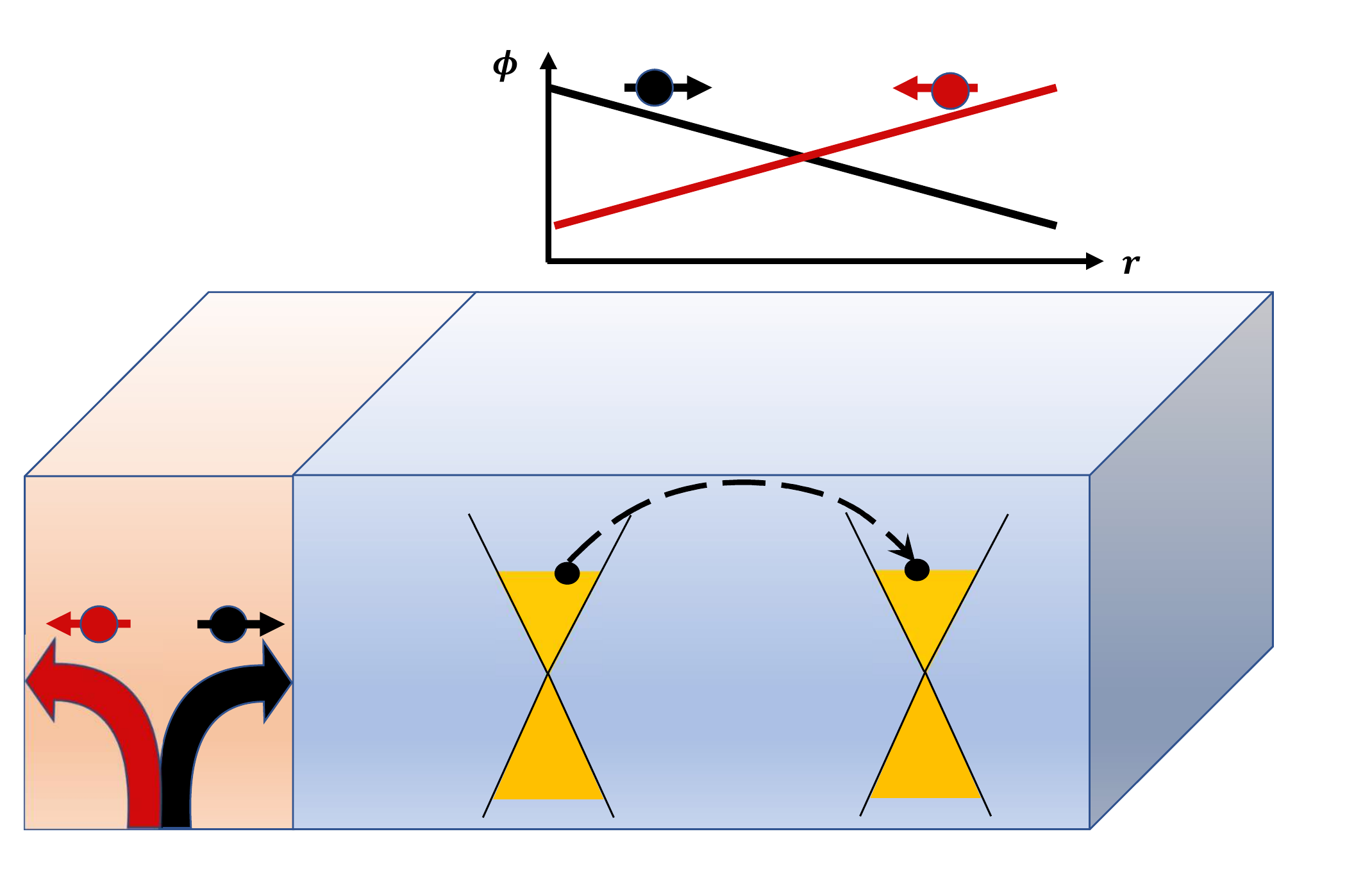}\\
  \caption{The schematic diagram of the chiral charge pumping from the spin current injection. The left material is a spin current source and generates a spin current with parallel spin polarization and flow direction. The resulting Zeeman energy potential for opposite spins are shown above.}\label{fig_fig1}
\end{figure}

{\it Spin current injection and chiral anomaly.}---We first sketch the general theory of the electron motion under the spin current injection. As illustrated in Fig.~\ref{fig_fig1}, a spin current source is attached to the Wey semimtal, which injects a spin current flowing along the $z$ direction, with the spin polarization along the same direction, into the Weyl semimetal. Such a spin current can be generated by  low-symmetry materials~\cite{MacNeill2017,Liu2019,Nan2020}.

The injected spin current establishes a Zeeman energy gradient across the Weyl semimetal. For example, for electrons with $s_z>0$ and $s_z<0$, the Zeeman energy is possitive and negative, and the driving force is along and against the injection direction, respectively, as shown in Fig.~\ref{fig_fig1}. Such an inhomgeneous Zeeman energy is amount to a Zeeman field $B_z^Z$ changing along the $z$ direction~\cite{Zelse2002,Shi2006}.

The inhomogeneous Zeeman field modifies the electron distrubtion through the following semiclassical Boltzmann equation
\begin{align}\label{eq_be}
\frac{\partial f}{\partial \bm k}\cdot \dot{\bm k}+\frac{\partial f}{\partial \bm r}\cdot \dot{\bm r}=\mathcal{I}_\text{collision}\,,
\end{align}
where $f$ is the electron distribution function and $\mathcal{I}_\text{collision}$ is the collision integral. 

At leading order, the distribution function $f$ on the left hand side can be replaced by the equilibrium Fermi distribution $f_0$. It is important to notice that since the inhomogeneous Zeeman energy is maintained by the spin current injection, the electron can only reach steady state but not equilibrium with it. This is similar to the electron dynamics under the electric field, where the electron cannot reach equilibrium with the electric potential. As a result, the inhomogeneous Zeeman energy cannot enter the distribution function $f_0$, i.e., $\partial f_0/\partial \bm r =0$.

To evaluate the first term in Eq.~\eqref{eq_be}, the force equation is needed. At the order of $\partial_zB_z^Z$, only the Zeeman energy gradient is relevant~\cite{suppl}, i.e., $\dot{ k}_z=-\partial_{z}\phi$ with  $\phi=g_S B_z^{Z} \langle \hat{s}_z\rangle$ being the Zeeman energy and $\langle \hat{s}_z\rangle$ being the expecation value of the spin. We have set $e=\hbar=\mu_B=1$ for simplicity. 
Plugging such force equation into Eq.~\eqref{eq_be}, we find that the left hand side has the following form
\begin{align}\label{eq_lhs}
\delta f=\frac{\partial f_0}{\partial \bm k}\cdot \dot{\bm k}=-g_S\langle \hat{s}_z\rangle v_z  \partial_zB_z^Z\frac{\partial f_0}{\partial \varepsilon}\,,
\end{align}
where $v_z=\partial_{k_z}\varepsilon$ is the band velocity.

Strikingly, Eq.~\eqref{eq_lhs} leads to the chiral anomaly in Weyl semimetals. To see this, we consider a Weyl node with the following Hamiltonian
\begin{align}\label{eq_ham}
\hat{H}^{(1)}&= v \bm k\cdot \bm \sigma\,,
\end{align}
where $v$ is the Fermi velocity and $\bm \sigma$ is the Pauli matrix in the pseudospin space. 

To evaluate $\delta f$, the spin expectation value is required. However, $\bm \sigma$ in Eq.~\eqref{eq_ham} does not in general operates in the real spin space directly. In fact, the spin information of the Weyl Fermion is hidden in the basis for the Weyl Hamiltonian. Label the eigenstate of $\sigma_z$ by $|+\rangle$ and $|-\rangle$, for the conduction band of the Weyl Hamiltonian, we have~\cite{suppl}
\begin{align}\label{eq_sj}
\langle \hat{s}_z \rangle_c&=\frac{\varepsilon+ vk_z}{2\varepsilon}\langle +|\hat{s}_z|+\rangle+\frac{\varepsilon- vk_z}{2\varepsilon}\langle -|\hat{s}_z|-\rangle\notag\\
&+\left( \frac{ v(k_x-ik_y)}{2\varepsilon}\langle +|\hat{s}_z|-\rangle+c.c.\right)\,.
\end{align}

The spin expectation value in Eq.~\eqref{eq_sj} contains a part that is proportional to $k_z$. Since $v_z\propto k_z$, we find that such a part of $\langle \hat{s}_z\rangle$ is always in-phase or out-of-phase with the velocity, depending on the sign of the expecation value of $\hat{s}_z$. This reflects the spin-momentum locking around a Weyl node.

The pumping rate of the partical number, i.e., $\delta P^{(1)}$ is obtained by the integration of $\delta f$ and hence determined by the flux of the spin field along the $z$-th direction acording to Eq.~\eqref{eq_sj}.
The above pattern of $\langle s_z\rangle$ can then enable a nontrivial $\delta P^{(1)}$. The final result reads
\begin{align}
\delta P^{(1)}&=-\frac{\mu^2}{12\pi^2 v^2} (\langle +|\hat{s}_z|+\rangle-\langle -|\hat{s}_z|-\rangle)g_S \partial_z B_z^Z\,.
\end{align}
where $\mu$ is the Fermi energy. One immediately finds that $\delta P^{(1)}$ is generally nonzero.

The pumping rate of particle number around the conjugate Weyl point can be obtained from symmetry analysis. We first consider the magnetic Weyl semimetal hosting a pair of Weyl nodes with opposite chiralities connected by the inversion symmetry. In Eq.~\eqref{eq_lhs}, $\langle \bm s\rangle$ and $\bm v$ transform as an axial and polar vector, respectively. Then under the inversion operation, $\langle \bm s\rangle\rightarrow \langle \bm s\rangle$ and $\bm v\rightarrow -\bm v$. As a result, the pumping rate of particle number around the coujugate Weyl point satisfies 
\begin{equation}
\delta P^{(2)}=-\delta P^{(1)}\,.
\end{equation}

Such opposite pumping rate of particle number is the essence of the chiral anomaly. It suggests that by injecting a spin current, the Weyl Fermion is always pumped out of a Weyl cone and into its conjugate partner. Originally, the inversion symmetry implies the degeneracy of the Fermi energy, i.e., $\mu^{(1)}=\mu^{(2)}=\mu$. The pumping process then lifts such degeneracy. By balancing the pumping process with the inter-Weyl-node scattering~\cite{Son2013}, we have
\begin{align}
\delta P^{(1)}-\delta P^{(2)}=-\frac{N^{(1)}-N^{(2)}}{\tau_i}\,,
\end{align}
where $\tau_i$ is the inter-Weyl-node relaxation time. Therefore, at the steady state, we have~\cite{suppl}
\begin{align}\label{eq_ca}
\mu^{(1)}-\mu^{(2)}=-\frac{v\tau_i}{3}(\langle +|\hat{s}_z|+\rangle-\langle -|\hat{s}_z|-\rangle)g_S \partial_z B_z^Z\,.
\end{align}

\begin{figure*}[t]
  \centering
  \includegraphics[width=0.95\linewidth]{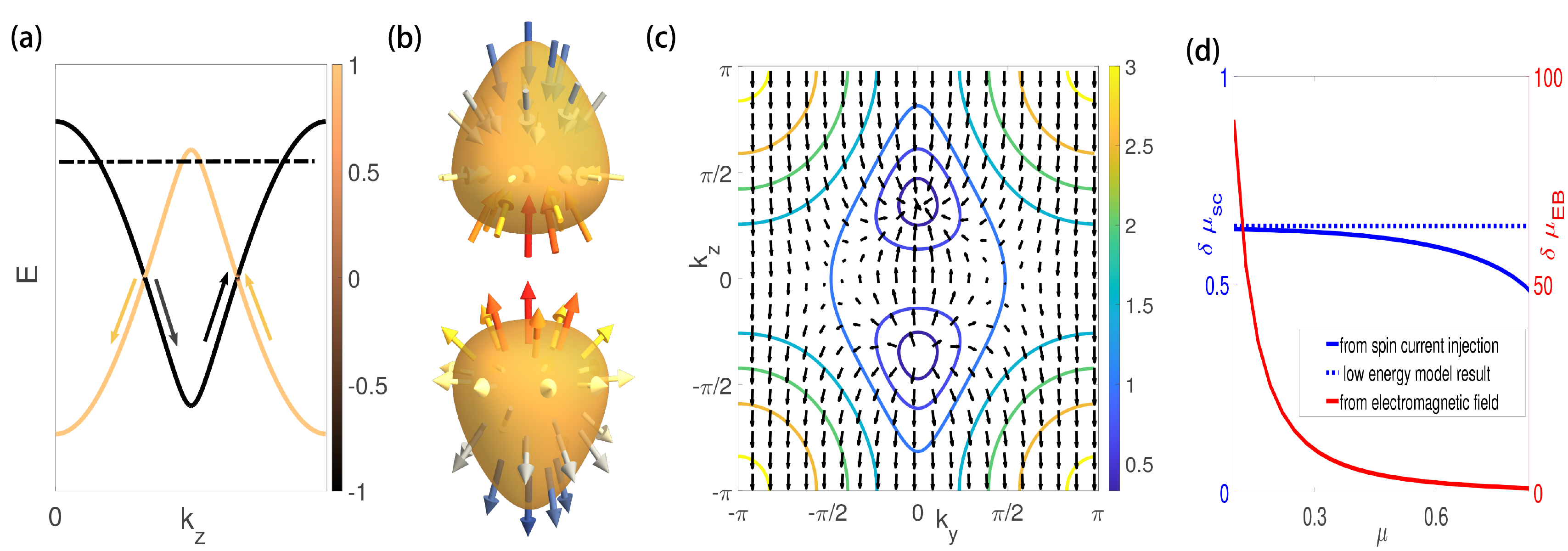}\\
  \caption{The topological origin of the chiral anomaly from spin current injection: (a) the band structure and spin projection; (b) spin field on a fixed equal-energy surface; (c) the spin field on the $k_y-k_z$ plane stacked by the equal-energy contour for the upper band of the Dirac cone; (d) the change in chemical potential due to spin current injection~(solid and dashed blue curves) and parallel electric and magnetic field~(red curve), respectively. In (a), beige and black colors represent up and down spin, and the corresponding states move along $-k_z$ and $k_z$ direction, respectively. In (d), the solid and dashed blue curve show the value of $\delta\mu$ for the full lattice model and the effective two-band Weyl Hamiltonian, respectively; the position of the maximum Fermi level is shown in the dashed curve in (a). }\label{fig_fig2}
\end{figure*}

The chiral polarization in Eq.~\eqref{eq_ca} is the main result of this work. To illuminate its physical meaning, we first note that, such chiral polarization is indeed a type of chiral anomaly. On one hand, the chiral charge density operator can be put in the form of $\sigma_0\otimes\tau_z$ with $\sigma_0$ being the identity matrix in the space spanned by $|+\rangle$ and $|-\rangle$ and $\tau_z=\pm 1$ labeling a pair of Wey nodes. On the other hand,
by maintaining the inhomgeneous Zeeman energy, we add a perturbative Hamiltonian of the form $H^\prime=g_S B_z^Z \hat{s}_z$ to the crystal Hamiltonian. We then project the perturbation onto the Hilbert space of the Weyl Hamitonian spanned by $|+\rangle$ and $|-\rangle$ and obtain 
\begin{equation}
\hat{H}_{proj}^\prime=g_SB_z^Z 
\begin{pmatrix}
 \langle +|\hat{s}_z|+\rangle & \langle +|\hat{s}_z|-\rangle\\ \langle -|\hat{s}_z|+\rangle & \langle -|\hat{s}_z|-\rangle 
\end{pmatrix}\,.
\end{equation}
The inversion symmetry then yields the perturbation Hamiltonian around the conjugate Weyl point, which is of the same form. Therefore, the perturbation can be put in the form of $\hat{H}_{proj}^\prime\otimes \tau_0$, with $\tau_0$ being the identity matrix within the same space of $\tau_z$. It is then clear that $\sigma_0\otimes\tau_z$ commutes with $\hat{H}_{proj}^\prime\otimes \tau_0$, implying the conservation of the chiral charge density. Eq.~\eqref{eq_ca} unexpectedly breaks such conservation of chiral symmetry, and hence introduces the chiral anomaly.

Secondly, the chiral polarization is consistent with that from parallel electric and magnetic field from the point-group symmetry perspective. Since the electric field $\bm E$ and magnetic field $\bm B$ are polar and axial vectors respectively, we find that $\bm E\cdot \bm B$ breaks the inversion symmetry and any type of mirror symmetry. In comprarison, the Zeeman field $B_z^Z$ is an axial vector and by taking spatial derivative, we introducing a polar vector component. As a result, $\partial_zB_z^Z$ has exactly the same symmetry requirements with $\bm E\cdot \bm B$. The chiral anomaly shown in Eq.~\eqref{eq_ca} thus should appear naturally.

However, the major difference in mechanism between the chiral polarization here and that in the real or strain-induced electromagnetic field case is that they rely on different aspects of the Weyl cone. The latter is built on the chiral flux from the Berry curvature, while the former does not require the participance of the Berry curvature at all. In fact, the chiral flux does not fully determine the spin field flux: the additional spin matrix element should be taken into account. Only in the simplest case of the Kramers-Weyl node~\cite{Chang2018}, where $|+\rangle=|\uparrow\rangle$ and $|-\rangle=|\downarrow\rangle$, the pseudospin operator coincides with the real spin operator in Eq.~\eqref{eq_ham}, and hence the spin flux is fully implied by the chiral flux.

Strikingly, Eq.~\eqref{eq_ca} suggests that the change in chemical potential is irrelevant of the position of the original chemical potential. In sharp contrast, the chemical potential difference from parallel electric and magnetic field is inversely proportional to $\mu^2$. As a result, the latter decreases fast when the Fermi energy moves away from the Weyl point, while the former stays the same.

The same result also exists in non-centrosymmetric Weyl semimetals. In this case, the Weyl nodes connected by the time reversal symmetry must have the same chirality. Therefore, a pair of Weyl nodes with opposite chiralities may be connected by other chiral operations, such as the mirror operation~\cite{Chan2016,Armitage2018}. Since $\langle \hat{s}_z\rangle v_z$ flips sign under any mirror operation, the pumping rates of electrons around conjugate Weyl points are opposite and the chiral polarization is hence still present.

{\it Topological origin.}---To reveal the topological origin of the chiral anomaly from the spin current injection, we define the following coefficient: $\delta P^{(1)}=-g_S \gamma_{ij} \partial_iB_j^Z$. Then we have
\begin{align}\label{eq_beta}
{\rm Tr}\gamma_{ij}=\int \bm v\cdot \langle\bm s\rangle f_0^\prime \frac{d\bm k}{8\pi^3}\,.
\end{align}
This implies that the variation of the particle number around a Weyl node is fully connected to the flux of the spin field. Therefore, the chiral anomaly exists as long as each Weyl cone is also a source or drain of the spin vector field. This is generally true, as $\gamma$ can be evaluted directly, and the result reads
\begin{align}
{\rm Tr}\gamma_{ij}=\frac{\mu^2}{12\pi^2 v^2} [\langle +|\hat{s}_z|+\rangle-\langle -|\hat{s}_z|-\rangle+(\langle+|\hat{s}_+|-\rangle+c.c.)]\,,
\end{align}
where $\hat{s}_+=\hat{s}_x-i\hat{s}_y$. We find that it is zero only when $\mu=0$ or the spin matrix element for the basis of the Weyl Hamiltonian satisfies a special condition.

To illustrate the topological origin of such chiral anomaly, we consider a concrete example containing a Weyl semimetal reguliarized on a cubic lattice~\cite{Koshino2016}, with the Hamiltonian given by
\begin{align}\label{eq_latham}
\hat{H}=\lambda \tau_x (s_x \sin k_x+s_y \sin k_y +s_z \sin k_z)+\epsilon\tau_z+b s_z\,,
\end{align}
where Pauli matrices $\bm \tau$ and $\bm s$ operates in orbital and spin space, respectively, $\epsilon=m+r(3-\cos k_x-\cos k_y-\cos k_z)$, $b$ is the Zeeman field, $\lambda$ is the strength of the spin-orbital coupling, and $\bm k$ is the dimensionless momentum. This model hosts two Weyl nodes connected by the inversion symmetry under appropriate choice of parameters, as shown in Fig.~\ref{fig_fig2}(a).

To show the motion of the electronic state under the spin current injection, we project the spin polarization of each state onto the $z$ direction along the $k_z$ axis in Fig.~\ref{fig_fig2}(a). Interestingly, for the each of the two bands composing the Weyl node, the spin always swithces after acrossing a Weyl point. This bahavior dictates that the electronic state is pumped from a Weyl node to the other one along the $k_z$ axis.

Such behavior extends to three dimension. As shown in Fig.~\ref{fig_fig2}(b), we plot the spin field on a fixed Fermi surface. Remarkably, the spin field approximately pointing outward on one surface and inward on the other one, allowing the inter-Weyl-node movement of the electronic state.

To further explore the extend to which such picture holds, we plot the spin field on the $k_y-k_z$ plane for the upper part of the Weyl cone, stacked by the equal-energy surface in Fig.~\ref{fig_fig2}(c). Since the rotational symmetry about the $k_z$ axis approximately holds around each Weyl node, such figure can well represent the general pattern of the spin field. The light blue curve shows where the two Weyl nodes merge. Inside it,  the distribution of the spin field resembles that of the elecric field from opposite electric charge density, demonstrating explicitly that as long as the Weyl node is well-defined, it can work as a source or drain of the spin field. Outside the merging curve, the spin distribution is dominated by the Zeeman effect.

As discussed previously, the source or drain nature of the spin field for each Weyl node is robust against the lattice reguralization of the Weyl point, protecting the chiral anomaly from the spin injection. However, the value of chiral polarization is not protected. According to Eq.~\eqref{eq_ca}, the change in chemical potential should be a constant of the Fermi energy. In Fig.~\ref{fig_fig2}(d), we plot such chemical potential change for the lattice model, represented by the blue curve. As a reference, we also plot $\delta \mu_{EB}$ from parallel electric and magnetic field, which decays as $1/\mu^2$ as the Fermi energy increases. We find that throughout the region where the Weyl cone is well defined, $\delta \mu_{SC}$ varies but is still approximately around a constant, i.e., the lattice regularization of the Weyl cone perturbs the leading order contribution to $\delta \mu_{SC}$ from the low-energy Weyl Hamiltonian.

{\it Experimental measurement.}---The chiral anomaly induced by the spin current injection can be probed experimentally in a similar way as that from the electromagnetic field. Specifically, by applying an additional magnetic field, it can generate a transport current. According to the chiral magnetic effect, for Weyl semimetals, a magnetic field can induce a following current\cite{Fukushima2008,Son2013,Zhou2013,Vazifeh2013,Burkov2015,Ma2015,Zhong2016}:
\begin{align}
\bm j=\frac{e^2}{4\pi^2 \hbar^2}\bm B\sum_i k^{(i)}\mu^{(i)}\,,
\end{align}
where $k^{(i)}$ is the chirality for the $i$-th Weyl node and $\mu^{(i)}$ is the Fermi energy. It has been shwon that the chiral magnetic effect can lead to a net current in the non-equilibrium setting~\cite{Armitage2018}. For example, if the chemical potential difference is maintained by an external electric and magnetic field as in the previously studied chiral anomaly, the chiral magnetic current can manifest as a large negative magnetoresistance~\cite{Son2013}.

Similarly the chiral anomaly from the spin current injection can also induce a chiral magnetic current. Since the magnetic field does not participate in the generation of the chemical potential difference, the current direction can be freely manipulated by simply rotating the direction of the magnetic field. Moreover, its magnitude approximately stays the same as the doping level varies. Both features are in sharp contrast to the chiral magnetic current from parallel electric and magnetic field, whose magnitude changes drastically with either the direction of the magnetic field or the doping level.

A possible optical signal to probe such chiral anomaly is the circular photogalvanic effect. It manifests as a photocurrent along the propagation direction of a circularly polarized light~\cite{Sipe2000}, i.e.,
\begin{align}
\frac{dJ_i}{dt}=\beta_{ij}(i\bm E\times \bm E^\star)_j\,,
\end{align}
with $i=j$. Such a current only exists in a material with structural chirality, i.e., both inversion symmetry and any mirror symmetry are broken.

In a Weyl semimetal with an achiral structure, such as those in the TaAs family with a mirror symmetry or centro-symmetric magnetic Weyl semimetal~\cite{Armitage2018}, $\beta_{ii}=0$ initially. In such a Weyl semimetal, a pair of Weyl nodes with opposite chiralities should be connected by either the inversion symmetry or the mirror symmetry. We further assume that the Weyl node has a nonzero Fermi energy $\mu>0$. It has been shown that although each Weyl node has a nonzero and constant contribution to $\beta_{ii}$, contributions from a pair of Weyl nodes always cancel~\cite{Juan2017}. 

The chiral anomaly here can remove such cancellation as it lifts the degeneracy in the Fermi energy. Then for the incident light with the frequency $2\omega=\mu$, the optical transition is forbidden in one Weyl node and allowed in the other one. Therefore, a nonzero $\beta_{ii}$ emerges. As the frequency futher increases or decreases such that both Weyl nodes contribution, the cancellation is recovered. Experimentally, this corresponds to a sharp peak in the optical current $\bm J$ at the frequency $2\omega=\mu$.

\begin{acknowledgments}
Y.G. acknowledges insightful discussions with Zhenyu Zhang, Changgan Zeng, Chong Wang, and Di Xiao, and support from the Startup Foundation from University of Science and Technology of China.
% D.X.\ also acknowledges the support of a Simons Foundation Fellowship in Theoretical Physics.
\end{acknowledgments}

\end{document}